\documentclass[twocolumn,prd,unsortedaddress,superscriptaddress,showpacs,a4paper,nofootinbib]{revtex4-1}
\usepackage[T1]{fontenc}
\usepackage{graphicx}
\usepackage{indentfirst}
\usepackage{pdfpages}
\usepackage{multirow}
\usepackage{amssymb}
\usepackage{amsfonts}
\usepackage{amsmath}
\usepackage{epsfig}

\def\beq{\begin{equation}}
\def\enq{\end{equation}}
\def\beqa{\begin{eqnarray}}
\def\enqa{\end{eqnarray}}
\def\MeV{\nobreak\,\mbox{MeV}}
\def\GeV{\nobreak\,\mbox{GeV}}

\def\qq{\lag\bar{q}q\rag}

\def\mix{\lag\bar{q}g\si.Gq\rag}
\def\mixs{\lag\bar{s}g\si.Gs\rag}
\def\Gd{\lag g^2G^2\rag}
\def\G3{\lag g^3G^3\rag}

\def\La{\Lambda}

\def\rh{\rho}
\def\si{\sigma}

\def\al{\alpha}

\def\lb{\label}
\def\nn{\nonumber}
\newcommand{\rag}{\rangle}
\newcommand{\lag}{\langle}

\begin{document}

\title{\sc A QCD sum rule study for a charged bottom-strange scalar meson}
\author{ C.M. Zanetti$^1$, M. Nielsen$^2$ and K.P. Khemchandani}
\affiliation{Faculdade de Tecnologia, Universidade do Estado do Rio de 
Janeiro, Rod. Presidente Dutra Km 298, P\'olo Industrial, 27537-000 , 
Resende, RJ, Brasil\\
$^2$Instituto de F\'{\i}sica, Universidade de S\~{a}o Paulo, 
C.P. 66318, 05389-970 S\~{a}o Paulo, SP, Brazil}

\begin{abstract}
Using  the QCD  sum rule approach  we investigate the  possible  four-quark   
structure  for the new  observed $B_{s}^0\pi^\pm$  narrow structure
(D0). We  use a diquak-antidiquark scalar current and work to the order of $m_s$ 
in full QCD, without relying  on $1/m_Q$  expansion.  Our  study indicates that although  it is possible to obtain a stable mass in agreement with 
the state found by the D0 collaboration, a more constraint analysis (simultaneous requirement of the OPE convergence and the dominance of the pole on
the phenomenological side) leads to a higher mass.
 We also predict the masses of the 
bottom scalar tetraquark resonances with zero and two strange quarks.
\end{abstract} 

\maketitle

Recently the D0 Collaboration reported the observation of
 a narrow structure, called $X(5568)$, in the decay $X(5568)\to B_{s}^0\pi^\pm$ 
\cite{D0:2016mwd}. 
This is the first observation of a hadronic state with two quarks and two 
antiquarks of four different flavors and, therefore, can only be explained as 
a tetraquark or molecular state. The mass and width of the observed state
were reported to be: $m=5567.8\pm2.9 (\mbox{sta})^{+0.9}_{-1.9}(\mbox{syst})
\MeV/c^2$ and $\Gamma=21.9\pm6.4 (\mbox{sta})^{+5.0}_{-2.5}(\mbox{syst})
\MeV/c^2$. As pointed out in Ref.~\cite{D0:2016mwd}, considering the large 
mass difference between the mass of the $X(5568)$ and the sum of the $B^0$ and 
$K^\pm$ masses, it can be difficult to explain the $X(5568)$ as a molecular
state. Therefore, the $X(5568)$ is an excellent candidate for a tetraquark 
state. If the $B_{s}^0\pi^\pm$ pair in the $X(5568)$ decay is produced in 
$S$-wave, its quantum numbers are $J^{P}=0^{+}$ as the very narrow 
$D_{s0}^+(2317)$ state, first discovered in the $D_s^+\pi^0$ decay channel by 
the BABAR Collaboration \cite{babar}. Due to its low mass, the structure of 
the $D_{s0}^{\ast\pm}(2317)$ meson has been extensively debated. It has been 
interpreted as a $c\bar{s}$ state \cite{dhlz,bali,ukqcd,ht,nari}, two-meson 
molecular state \cite{bcl,szc,Kolomeitsev:2003ac,Guo:2006fu,gamermann,Guo:2009ct,Cleven:2010aw,Cleven:2014oka,Faessler:2007cu,Faessler:2007gv}, 
$K-D$- mixing \cite{br},  four-quark states \cite{ch,tera,mppr,nos} or a 
mixture between two-meson and four-quark states \cite{bpp}. 
In this paper we use the QCD sum rule (QCDSR) approach 
\cite{svz,rry,SNB,Nielsen:2009uh,Nielsen:2014mva}  to investigate the 
possible four-quark structure for the $X(5568)$ and, therefore, to test if
the $X(5568)$ could be the isovector bottom partner of the  $D_{s0}^{+}(2317)$.

The QCDSR for  scalar mesons are constructed from the two-point
correlation function written in terms of a scalar current $j_S$:
\beq
\Pi(q)=i\int d^4x ~e^{iq.x}\lag 0 |T[j_S(x)j^\dagger_S(0)]|0\rag.
\lb{2po}
\enq

The key idea of the QCDSR method is to consider that this correlation 
function is of dual nature and it depends on the value of the momentum $q$. 
For large momentum, i.e., short distances, the correlation function can be 
calculated using perturbative QCD. In this case, the current $j_S$ is written 
in terms of the quark content of the studied mesons. However, since we are 
interested in studying the properties of hadrons, the relevant energies are 
lower and contributions from quark condensates, gluon condensates, etc., need 
to be included in the evaluation of Eq.~(\ref{2po}).  This can be done by 
using the Wilson operator product expansion (OPE) of the correlation 
function. In this case,  Eq.~(\ref{2po}) is expanded in terms 
of local condensates and a series of coefficients. The local operators 
incorporate nonperturbative long-distance effects, while the coefficients, 
by construction, include only the short-distance domain and can be determined 
perturbatively. This way of evaluating the correlation function is customarily named as
the calculation on the 
``OPE side".  

At large distances, or, equivalently, small momentum, the currents 
$j^\dagger_S$ and $j_S$ of Eq.~(\ref{2po}) can be interpreted as operators of 
creation and annihilation of the scalar mesons. In this case, the correlation 
function is obtained by inserting a complete set of scalar states.
This interpretation of the correlation function is called as the 
``phenomenological side". The assumption made in the QCDSR approach is that 
there must be a range of $q^2$ values in which both descriptions must be 
equivalent. Calculating the correlation function of Eq.~(\ref{2po})  using 
these two approaches and equating them, it is possible to obtain information 
about the properties of the hadronic states generated in the system.

In Ref.~\cite{nos} the  $D_{s0}^{+}(2317)$ state was considered as a 
diquark-antidiquark tetraquark state and was studied by using the QCDSR 
approach. A very good agreement with the experimental mass was obtained. 
Here we follow Ref.~\cite{nos} to write an analogous but isovector 
scalar-diquark scalar-antidiquark tetraquark current for $X(5568)$:
\beqa
j_S&=&{\epsilon_{abc}\epsilon_{dec}}(u_a^TC
\gamma_5s_b)(\bar{d}_d\gamma_5C\bar{b}_e^T),
\label{int}
\enqa
where $a,~b,~c,~...$ are colour indices, $C$ is the charge conjugation
matrix. 
 Of course a scalar-scalar diquark-antidiquark  form is not the only possible choice for a 
scalar tetraquark
current, and one could use pseudoscalar-pseudoscalar, vector-vector or axial-axial
diquark-antidiquark form. However, it was shown in Ref.~\cite{Kim:2005gt} that the 
scalar-scalar type of current gives the more stable results. Therefore, we use the current given by Eq.~(\ref{int}) for the $X^+(5568)$.

The coupling of the state, $X$, to the scalar current, $j_S$,  can be
parametrized in terms of the  constant $f_X$ as:
$\lag 0 | j_S|X\rag =f_X$,
therefore, the phenomenological side of Eq.~(\ref{2po}) can be written as
\beq
\Pi^{phen}(q^2)={f_X^2\over m_X^2-q^2}+\cdots\;,
\lb{phe}
\enq
where the dots denote the contribution from higher resonances,  which is usually 
parametrized through the introduction of a continuum threshold
parameter $s_0$ \cite{io1}.

On the OPE side we work at leading order and consider condensates up to 
dimension six. We deal with the strange quark as a light one and consider
the diagrams up to order $m_s$. To keep the bottom quark mass finite, we
use the momentum-space expression for the bottom quark propagator. We follow 
ref.~\cite{su} and calculate the light quark part of the correlation
function in the coordinate-space, which is then Fourier transformed to the
momentum space in $D$ dimensions. The resulting light-quark part is combined 
with the charm-quark part before it is dimensionally regularized at $D=4$.

We can write the correlation function on the OPE side in terms of a 
dispersion relation:
\beq
\Pi^{OPE}(q^2)=\int_{m_b^2}^\infty ds {\rho(s)\over s-q^2}\;,
\lb{ope}
\enq
where the spectral density is given by the imaginary part of the correlation
function: $\rho(s)={1\over\pi}\mbox{Im}[\Pi^{OPE}(s)]$. After making a Borel
transform on both sides, and transferring the continuum contribution to
the OPE side, the sum rule for the scalar meson $X$ can be written as
\beq
f_X^2e^{-m_X^2/M^2}=\int_{m_b^2}^{s_0}ds~ e^{-s/M^2}~\rho(s)\;,
\lb{sr}
\enq
where $M$ is the Borel mass and $\rho(s)=\rho^{pert}(s)+\rh^{m_s}(s)+\rh^{\qq}(s)+\rh^{\lag G^2\rag}
(s)+\rh^{mix}(s)+\rh^{\qq^2}(s)+\rh^{\lag G^3\rag}(s)$, with
\beq\label{rhopert}
\rho^{pert}(s)={1\over 2^{10} 3\pi^6}\int_\La^1 d\al\left({1-\al\over\al}
\right)^3(m_b^2-s\al)^4,
\enq
\beq\rh^{m_s}(s)=0,\enq
\beqa
\rho^{\qq}(s)={1\over 2^{6}\pi^4}\int_\La^1 d\al~{1-\al\over\al}
(m_b^2-s\al)^2
\bigg[
\nn\\
-\qq\left(2m_s+m_b{1-\al\over\al}\right)+m_s\lag\bar{s}s\rag\bigg],
\enqa
\beqa
\rho^{\lag G^2\rag}(s)&=&{\Gd\over 2^{10}\pi^6}\int_\La^1 d\al~(m_b^2-s\al)
\left[{m_b^2\over9}\left({1-\al\over\al}\right)^3+\right.
\nn\\&+&
\left.(m_b^2-s\al)\left({1-\al\over2\al}+{(1-\al)^2\over4\al^2}\right)
\right],\label{rhog2}
\enqa
\beqa
\rho^{mix}(s)&=&{1\over 2^{6}\pi^4}\int_\La^1 d\al~(m_b^2-s\al)\bigg[-{m_s\mixs
\over6}
\nn\\
&+&\mix\bigg(-m_s(1-\ln(1-\al))
\nn\\
&-&m_b{1-\al\over\al}\left(1-{1-\al\over2\al}\right)\bigg)
\bigg],
\enqa
\beqa
\rho^{\qq^2}(s)&=&-{1\over 24\pi^2}\int_\La^1 d\al~\bigg(2\qq\lag\bar{s}s\rag(m_b^2-s\al+m_bm_s)\nonumber\\&&-\qq^2m_bm_s\bigg),
\enqa
\beq
\rho^{\lag G^3\rag}(s)={\G3\over 2^{12} 9\pi^6}\int_\La^1 d\al\left({1-\al
\over\al}\right)^3(3m_b^2-s\al),\label{rhog3}
\enq
where the lower  limit of the integrations is  given by $\La=m_b^2/s$.

In order to  compute the mass of  the state, $m_X$, we  first take the
derivative of Eq.~(\ref{sr})  with respect to $1 /M^2$ and  then we divide the
result by Eq.~(\ref{sr}),  obtaining
\beq
m_X^2={\int_{m_b^2}^{s_0}ds ~e^{-s/M^2}~s~\rh(s)\over\int_{m_b^2}^{s_0}ds 
~e^{-s/M^2}~\rh(s)}\;.
\lb{m2}
\enq
This expression will be used to evaluate the mass of the state.

The numerical values for the quark masses and condensates are listed in
Table~\ref{tab1} \cite{SNB,snc,snd,sne}. 
\begin{center}
\begin{table}[h]
\begin{tabular}{p{3.5cm} p{3.5cm}}
\hline
Parameters & Values \\
\hline
 $m_s$ &  $(0.13\pm0.03)\,\GeV$ \\
 $m_b$ & $(4.24\pm0.06)\,\GeV$ \\
 $\lag\bar{q}q\rag$ &  $-(0.23\pm0.03)^3\,\GeV^3$ \\
$\langle\overline{s}s\rangle$ & $(0.8\pm0.2)\lag\bar{q}q\rag$ \\
 $m_0^2=\lag\bar{q}g\si.Gq\rag/\lag\bar{q}q\rag$  &  $0.8\,\GeV^2$ \\
 $\lag g^2G^2\rag$ &  $(0.88\pm0.25)~\GeV^4$\\
$\lag g^3G^3\rag$  &  $(0.58\pm0.18)~\GeV^6$ \\
\hline
\end{tabular}\caption{QCD input parameters.}\label{tab1}\end{table}
\end{center}

The remaining input to the calculation is the continuum threshold  parameter which, in general, is related to the mass of the state to be studied 
($X(5568)$, in the present case) as $s_0\sim(m_X+0.5~\GeV)^2$. Therefore, to start our analysis 
we choose $\sqrt{s_0} \sim6.0$ GeV.

\begin{figure}[t]
\centerline{\epsfig{figure=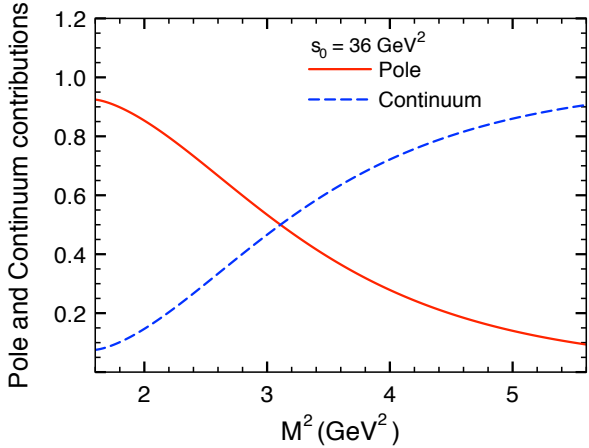,height=65mm}}
\caption{The pole (solid line) and the continuum (dashed line) contribution for 
$\sqrt{s_0}=6.0\GeV$.} \label{fig1}
\end{figure}

In  order to  determine the  values of  the Borel  mass parameter,  we
analyze  the pole  contribution, the OPE convergence and the Borel stability.   
In the  QCDSR  approach we  extract
information only from  the ground state, therefore we  must ensure that
the pole contribution is greater that the continuum contribution. Here
we fix the Borel mass in such a way that the pole contribution is always
between 80\% and   50\%  of   the  total  contribution.    From  
Fig.~\ref{fig1}  we can  see  that this  condition  is satisfied  for
values of  the  Borel mass  in the range $2.2\GeV^2\le  M^2\le3.0\GeV^2$.

\begin{figure}[h]
\centerline{\epsfig{figure=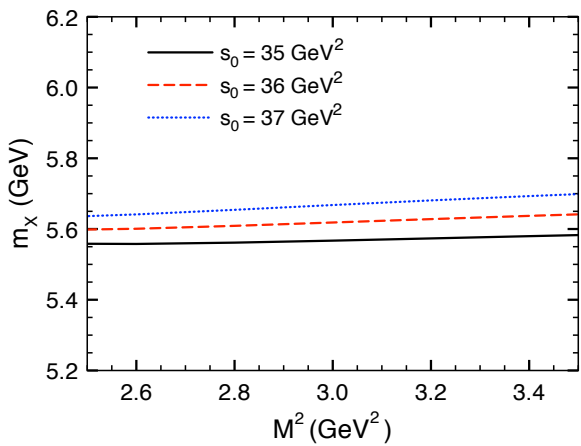,height=65mm}}
\caption{The  $X$ mass  as  a  function of  the  Borel mass  for
  different  values   of  the  continuum  threshold:  
  $\sqrt{s_0}=5.9\GeV$ (solid line);  $\sqrt{s_0}=6.0\GeV$ (dashed line); 
$\sqrt{s_0}=6.1\GeV$ (dotted line).} \label{fig2}
\end{figure}

In Fig.~\ref{fig2}  we plot  the ground  state mass  as a  function of
$M^2$,   considering  three   different   values   of  the   threshold
parameter. We  can see that  there is  a good $M^2$-stability  for the
Borel window considered. Using the central values of the parameters in 
Table~\ref{tab1} and $s_0= 36\;\GeV^2$ we get
\beq  m_X\sim5.58\;\GeV.   \enq

To evaluate the uncertainties inherent to the QCD sum rule approach, we  
consider the variation of the mass in  Borel  window, as a function of the  
continuum threshold, changed within a small range: $5.9\leq\sqrt{s_0}\leq6.1\GeV$, and the  quark masses and 
condensates errors indicated  in Table~\ref{tab1}.  Considering these uncertainties 
we get:
\beq m_X=(5.58\pm0.17)\GeV,   \label{result1}\enq 
 which  is  in excellent agreement  with  the
experimental   mass   of   the    X(5568)   determined   by   the   D0
Collaboration \cite{D0:2016mwd}.

\begin{figure}[t] 
\centerline{\epsfig{figure=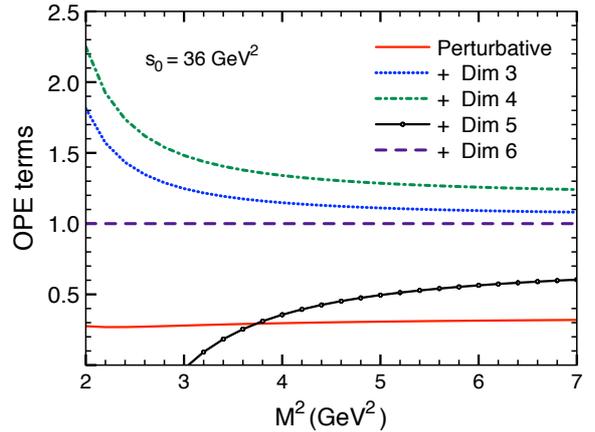,height=65mm}}
\caption{The OPE convergence in the region $2.0 \leq M^2 \leq
7~\GeV^2$ for ${s_0} = 36 \GeV$. We start with the relative perturbative
contribution (the perturbative contribution divided by the total contribution) and each
subsequent line represents the addition of the relative contribution of a 
condensate of higher dimension in the expansion.}
\label{fig3} 
\end{figure}

The result in Eq.~(\ref{result1}) was obtained considering only the pole dominance 
and the stability with the Borel mass. There is, however, a stronger constraint to 
the lower bound of the $M^2$, that comes from imposing the OPE convergence.
We  analyze the convergence of the OPE by comparing the relative 
contribution of each term  given by Eqs.~(\ref{rhopert}) to (\ref{rhog3}),
to the right hand side of Eq.~(\ref{sr}). The  requirement of a good
convergence sets a lower limit to $M^2$. This analysis in shown
in Fig.~\ref{fig3}.

As can be seen from Fig.~\ref{fig3}, there is no OPE convergence in any region allowed by the upper bound 
given by pole/continuum analysis: $M^2\leq3.0 \GeV^2$.
This means that the lower bound given by OPE convergence will be higher
than the upper bound, and there is no valid ``sum rule window'' where we can
completely trust the  results for this current.

To overcome this problem we can consider higher dimension condensates in the OPE side
and test if the series starts converging with such contributions. We include the condensates of dimension seven and eight whose expressions are
given below:
\beqa       
&&\rho^{\lag\bar{s}s\rag\lag        G^2\rag}(s)=\frac{m_s\lag\bar{s}s\rag\Gd}{2^93^2       
  \pi^4}\left[3\int_\Lambda^1d\al +\right.\nonumber\\&&\left.-\int_0^1d\al\frac{2m_b^2\al}{(1-\al)^2}\delta\left(s-\frac{m_b^2}{1-\al}\right)\right],
\enqa
\beqa
\rho^{\qq\lag G^2\rag}(s)&=&-\frac{\qq\Gd}{2^8 3^2 \pi^4}\left[ m_s+\int_\Lambda^1d\al\bigg((4m_b+3m_s)+\right.\nonumber
\enqa
\beqa
&&+\frac{3(1-\al)}{\al}\bigg(\frac{m_b(1-\al)}{\al}-3m_b\bigg)\bigg)\nonumber\\&&+\int_0^1d\al\,\frac{m_b^2\al}{(1-\al)^2}\bigg(\frac{1}{2}-\frac{m_b\al}{1-\al}\bigg)\delta\bigg(s-\frac{m_b^2}{1-\al}\bigg)\bigg],
\enqa

\beqa
&&\rho^{\qq\lag \bar{q}Gq\rag}(s)=\frac{m_bm_s\qq\mix}{2^43\pi^2}\bigg[\int_0^1d\al\frac{1}{1-\al}\times\nonumber\\&&\times\delta\bigg(s-\frac{m_b^2}{1-\al}\bigg)-2\,\delta(s-m_b^2)\bigg],
\enqa

\beqa
&&\rho^{\lag\bar{s}s\rag\lag \bar{q}Gq\rag}(s)=-\frac{\lag\bar{s}s\rag\mix}{2^53\pi^2}\bigg[2-m_bm_s\,\delta(s-m_b^2)+\nonumber\\&&+\int_0^1\, d\al\frac{m_bm_s}{1-\al}\delta\bigg(s-\frac{m_b^2}{1-\al}\bigg)\bigg],
\enqa

\beqa
\rho^{\qq\lag \bar{s}Gs\rag}(s)=-\frac{\qq\mixs}{2^53^2\pi^2}\bigg(6+\delta(s-m_b^2)\bigg),
\enqa

\beqa
\rho^{\lag g^4G^4\rag}(s)&=&-\frac{\lag g^4G^4\rag}{2^{14}3\pi^6}\bigg[\int_\Lambda^1d\al+\frac{2}{3}m_b^2\int_0^1d\al \frac{\al}{(1-\al)^2}\times\nonumber\\&&\times\delta\bigg(s-\frac{m_b^2}{1-\al}\bigg)\bigg].
\enqa

\begin{figure}[t] 
\centerline{\epsfig{figure=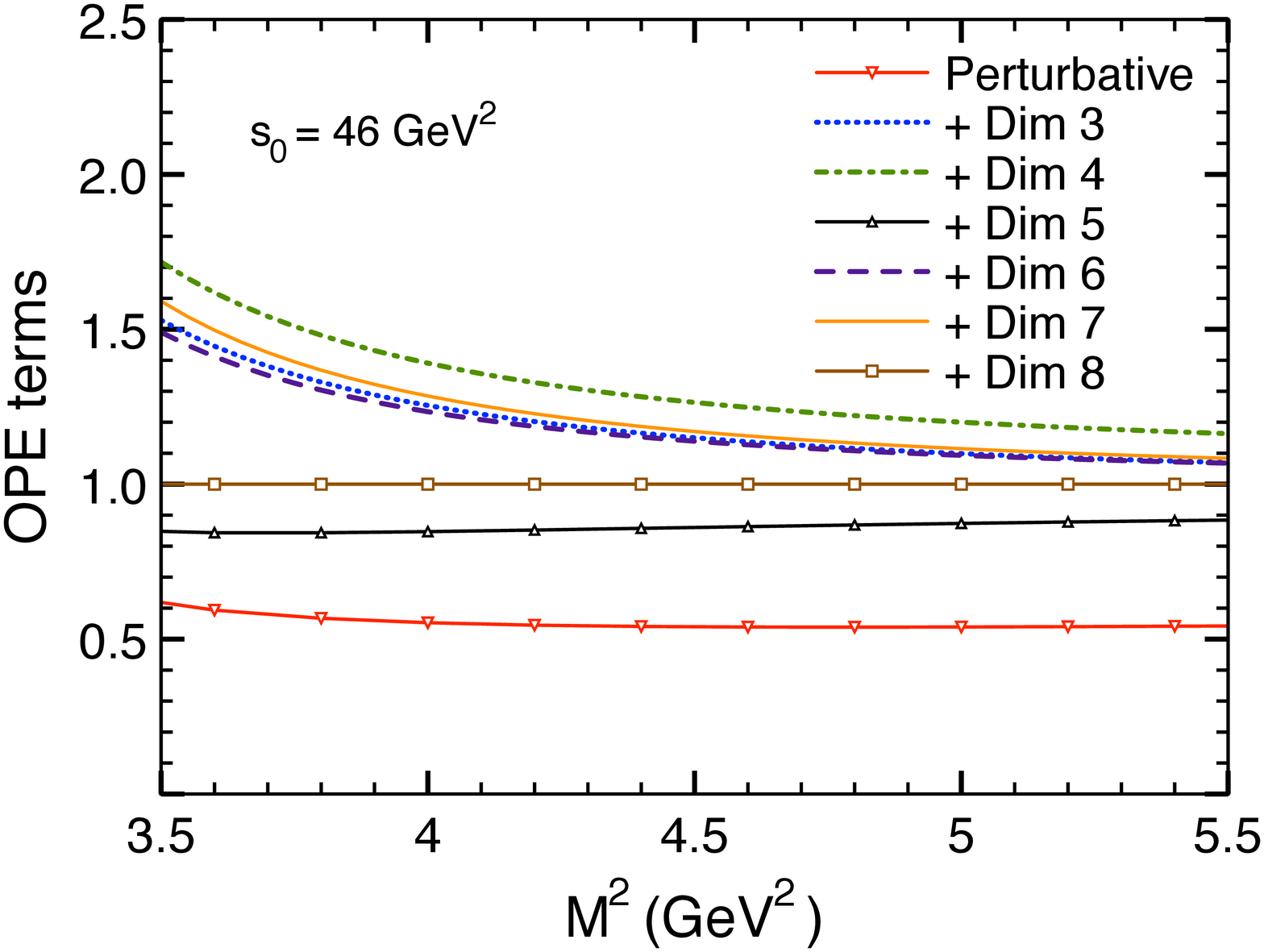,height=70mm}}
\caption{The OPE convergence in the region $3.5 \leq M^2 \leq5.5\GeV^2$
for ${s_0} = 46 \GeV$. We start with the relative perturbative
contribution (the perturbative contribution divided by the total contribution) and each
subsequent line represents the addition of the relative contribution of a 
condensate of higher dimension in the expansion.}
\label{fig4} 
\end{figure}

\begin{figure}[t]
\centerline{\epsfig{figure=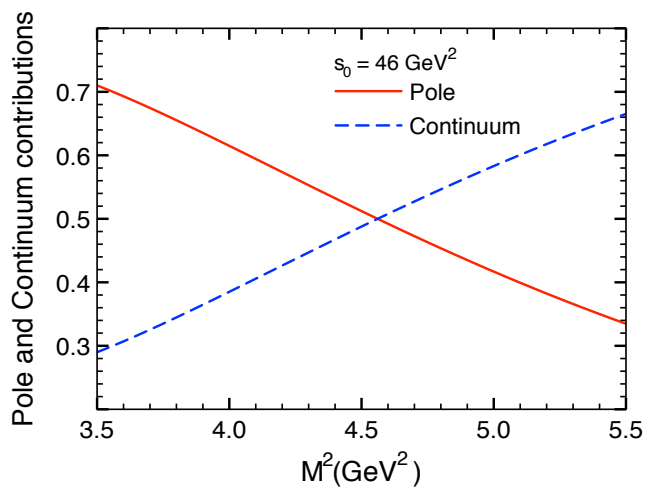,height=70mm}}
\caption{The pole (solid line) and the continuum (dashed line) contribution for 
$s_0=46.0\GeV^2$.} \label{fig5}
\end{figure}

On continuing with our analysis we find that, even after considering condensates up to dimension eight,  a valid ``sum rule window'' exists only
for values of $s_0\geq 46~\GeV^2$. In Figs.~\ref{fig4} and \ref{fig5} we show the
OPE convergence and the pole versus continuum contribution for  $s_0\geq 46~\GeV^2$.

From Fig.~\ref{fig4}  we can  see that there  is an OPE  convergence, the
dimension eight  condensate contribution is  smaller than 20\%  of the
total  contribution, only  for values  of $M^2\geq4.4~\GeV^2$.  On the
other hand, from Fig.~\ref{fig5} we can see that the pole contribution
is   bigger   than   the   continuum  contribution   for   values   of
$M^2\leq4.5~\GeV^2$. Although very  small, there exists a  valid Borel window in the region
$4.4\GeV^2\leq  M^2\leq4.5\GeV^2$, which  provides  a  good ``sum  rule''  to extract a reliable value  for the mass of
the state.

\begin{figure}[t]
\centerline{\epsfig{figure=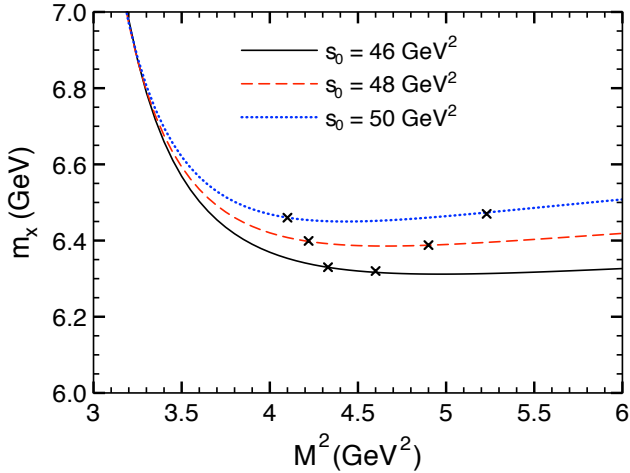,height=70mm}}
\caption{The  $X$ mass  as  a  function of  the  Borel mass  for
  different  values   of  the  continuum  threshold:  
  ${s_0}=46\GeV^2$ (solid line);  ${s_0}=48\GeV^2$ (dashed line); 
${s_0}=50\GeV^2$ (dotted line). The crosses in the figure indicate the allowed
Borel window.}
\label{fig6}
\end{figure}

In Fig.~\ref{fig6}, we show the resulting value for the mass of the state, as a
function of the Borel mass, for three different values of the continuum threshold.
The crosses in the figure indicate the ``sum rule window''.

Finally, considering all the restrictions described above we get:
\beq m_X=(6.39\pm0.10)\GeV,   \label{result2}\enq 
 which  is  not in agreement  with  the experimental   mass   of   the    $X(5568)$   
determined   by   the   D0  Collaboration \cite{D0:2016mwd}. As a matter of fact, 
recently the LHCb Collaboration has not confirmed the observation of the $X(5568)$.
In their preliminary analysis \cite{LHCb} no structure is found
in the $B_{s}^0\pi^\pm$ mass spectrum from the  $B_{s}^0\pi^+$ threshold up to 
$M_{B_s^0\pi^+}\leq 5700\GeV$. More  analyses are required to clarify this situation.
Our work predicts a  tetraquark state decaying in this channel  with a mass 
around 6.39 GeV. 

The uncertainty given in 
Eq.~(\ref{result2})  is only related with the  range of values of the Borel  mass 
window, a small variation in the  continuum threshold, $46\leq {s_0}\leq50\GeV^2$, and 
the  quark masses and condensates errors indicated  in Table~\ref{tab1}. The difference 
between the values in Eqs.~(\ref{result1}) and (\ref{result2}) can be associated 
mainly with the change in the value of the continuum threshold. However, as discussed
above, there is no allowed Borel window for values of $s_0\leq46~\GeV^2$. Therefore,
the result obtained with $s_0\sim36~\GeV^2$ given in Eq.~(\ref{result1}), although
being obtained in a Borel region where one has pole dominance, can not be trusted.
This result illustrates very well how we can reproduce the mass of a 
given state and then after a more careful analysis conclude that the state 
is not the particle associated with the chosen current. We also would like to point
out that the difference between $\sqrt{s_0}=\sqrt{48}~\GeV$ and the result in 
Eq.~(\ref{result2}) is about 0.5GeV, as the general supposition of the QCDSR approach for the 
start of the continuum threshold. It is also important to notice that the difference: 
$\sqrt{s_0}-m_X$ increases with the value of the continuum threshold. As an example, for
$s_0\sim64~\GeV^2$ we get $m_X\sim6.7~\GeV$ which implies $\sqrt{s_0}-m_X\sim1.3~\GeV$,
much larger than 0.5 GeV. This could be an indication that there is a contribution
from higher resonances below the continuum threshold and, therefore, once again, the estimated mass
can not be trusted. Therefore, to fix a ``good range'' of the values of $s_0$ we test if 
it  provides an allowed Borel window (where both constraints of the pole dominance and  the ope
convergence are satisfied), and that the 
 value of the obtained mass falls  within the range 0.4 GeV to 0.6 GeV smaller than $\sqrt{s_0}$.
Using these criteria, we have obtained $s_0$ in the range $46\leq {s_0} \leq50~ \GeV^2$.

We can extend the formalism to  study bottom scalar mesons states that
contain  zero and  two  strange  quarks.  In  order  to calculate  the
correlation   function  for   these  states   we  use   the  following
interpolating fields  for these states  (zero and two  strange quarks,
respectively):

\beqa
j_0&=&\epsilon_{abc}\epsilon_{dec}(u_a^TC\gamma_5q_b)
(\bar{d}_d\gamma_5C\bar{b}_e^T),
\nn\\
j_{ss}&=&\epsilon_{abc}\epsilon_{dec}(u_a^TC
\gamma_5s_b)(\bar{s}_d\gamma_5C\bar{b}_e^T),
\label{int2}
\enqa
where $q$ represents the quark $u$ or $d$ according to the charge of the meson.
The expression for the resulting spectral densities are given in Appendix~A.

\begin{figure}[t]
\centerline{\epsfig{figure=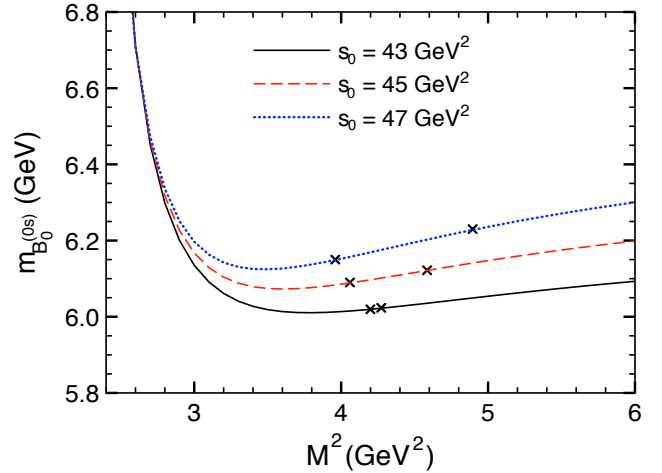,height=70mm}}
\caption{The $B_0^{0s}$ mass as a function of the Borel mass 
for different
values of the continuum threshold:  ${s_0}=43\GeV^2$ (solid line);
  ${s_0}=45\GeV^2$ (dashed line);   ${s_0}=47\GeV^2$ (dotted line). The crosses in the figure indicate the allowed
Borel window.} \label{fig7}
\end{figure}

\begin{figure}[t] 
\centerline{\epsfig{figure=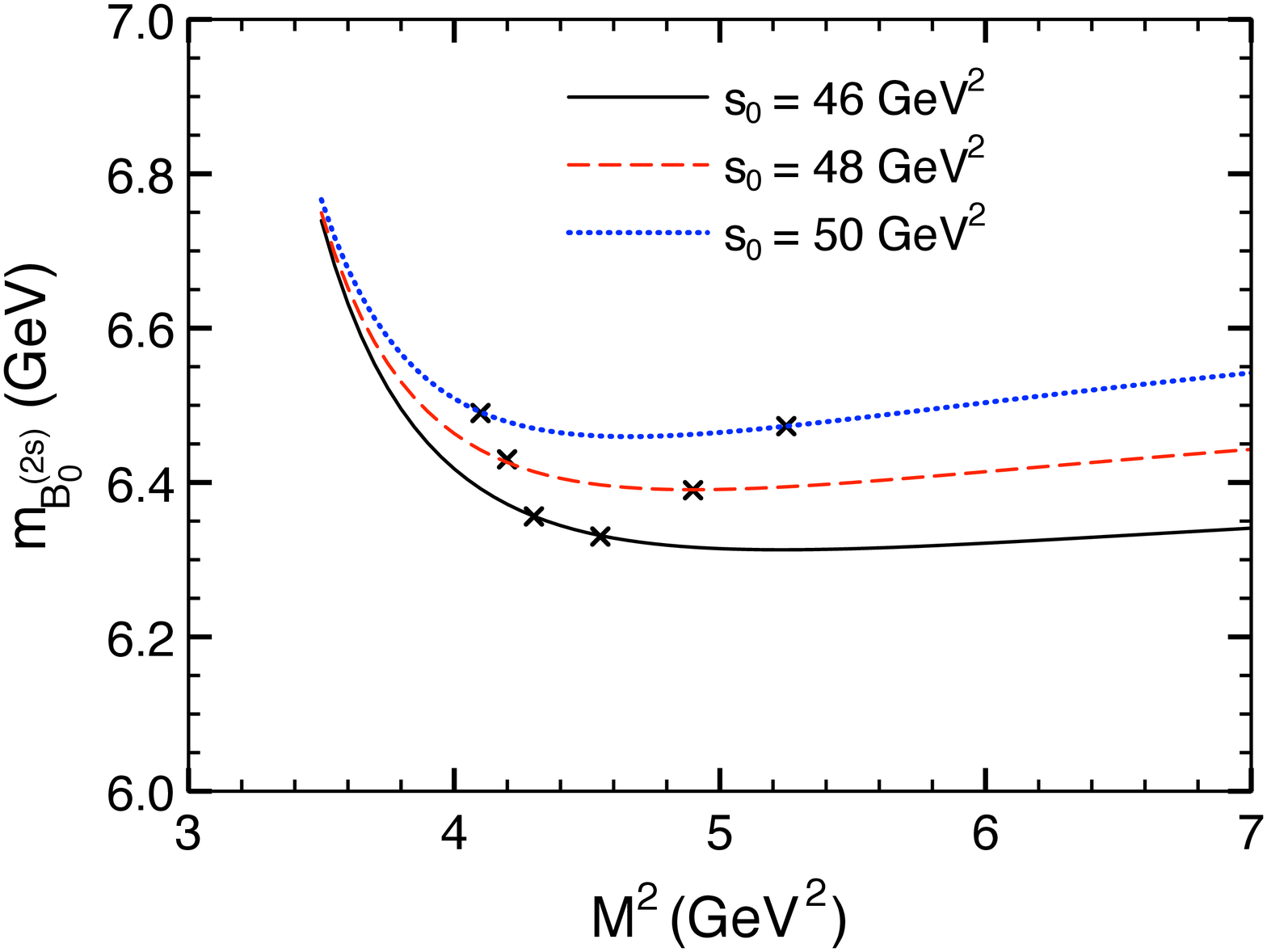,height=70mm}}
\caption{The $B_0^{2s}$ mass as a function of the Borel mass 
for different
values of the continuum threshold:  
  $\sqrt{s_0}=46\GeV^2$ (solid line);  ${s_0}=48\GeV^2$ (dashe line); ${s_0}=50\GeV^2$ (dotted line). The crosses in the figure indicate the allowed
Borel window.} \label{fig8}
\end{figure}

We  call  $B_0^{(0s)}$  and  $B_0^{(2s)}$  the  scalar  bottom  tetraquark mesons
represented by  $j_0$ and  $j_{ss}$ respectively  (Eq.~\ref{int2}). In
Figs.~\ref{fig7} and \ref{fig8} we show the  masses of the states as a
function  of the  Borel mass  for  different values  of the  continuum
threshold, and as  the previous case there is  good $M^2$-stability in the
allowed Borel window, represented by the crosses in these figures.  




The values obtained for the masses of the states are:
\beqa
m_{B_0^{0s}}=6.10\pm0.16\GeV,\\
m_{B_0^{2s}}=6.39\pm0.17\GeV.\label{result3}
\enqa
The sources of errors are the same that were used for the strange-bottom scalar meson.

Comparing      the       results      of      the       masses      in
Eqs.~(\ref{result2}) and (\ref{result3})  we can
see that  the $X$  (the state with one  strange quark)  and $B_{0}^{(2s)}$
resonance  masses  are  basically   degenerated,  while  the  mass  of
$B_{0}^{(0s)}$ is around $300\MeV$ smaller  than the others.  The same
behavior is observed for the scalar  mesons in the charm sector, as it
was pointed out in  Ref.~\cite{nos}.  The increase in the  mass is expected
with the  inclusion of  one strange  quark (from  zero to  one strange
quarks).  The fact this is not observed  when one goes from one to two
strange quarks can  be traced when we compare the  contribution of the
quark condensate  term, that  is smaller  for the  $B_{0}^{(2s)}$, but
this  is  compensated   by  the  inclusion  of   a  term  $\rho^{m_s}$
(Eq.~\ref{rhoms}).

In summary, we have presented a QCD sum rule study for a bottom scalar
meson considered as diquark-antidiquark state. The motivation of the study is
to look for a possible state associated with the recently claimed  $X(5568)$  by the D0  Collaboration.
We find that  it is possible to obtain a stable mass in agreement with 
the state found by the D0 collaboration while satisfying the condition of the pole
dominance on the phenomenological side, but by sacrificing  the simultaneous constraint of  the OPE convergence.
This last missing ingredient casts a doubt  on the reliability of the result, leading us to conclude  that  the  $X(5568)$  state
can not be represented by the scalar tetraquark current. 
We find that a rigorous application of QCD sum rule constraints leads to a higher mass.
Thus, we predict the existence of a scalar bottom-strange tetraquark state
with a mass around 6.4 GeV.  We also have obtained
the  masses for  the  resonances  with zero  and  two strange  quarks, which we call
$B_0^{0s}$ and  $B_0^{2s}$.  These  resonances have also not  been observed yet,
but  our calculations show that they should exist.

\vspace{0.5cm}

\underline{Acknowledgements}: 
This work has been supported by CNPq and FAPESP. 
\vspace{0.5cm}

\appendix
\label{apa}
\section{The spectral densities for resonances with zero and two strange quarks}

The expressions  (\ref{rhopert}), (\ref{rhog2}) and  (\ref{rhog3}) are
common to  the three resonances. Next,  we list the other terms
that are not common.

From $j_0$ we get: 
\beq\rh^{m_s}(s)=0,\enq
\beq
\rho^{\qq}(s)=-{m_b\qq\over 2^{6}\pi^4}\int_\La^1 d\al\left({1-\al\over\al}
\right)^2(m_b^2-s\al)^2,
\enq
\beqa
\rho^{mix}(s)&=&{m_b\mix\over 2^{6}\pi^4}\bigg[{1\over2}\int_\La^1 d\al
\left({1-\al\over\al}\right)^2(m_b^2-s\al)+
\nn\\
&-&\int_\La^1 d\al{1-\al\over\al}(m_b^2-s\al)\bigg],
\enqa
\beq
\rho^{\qq^2}(s)=-{\qq^2\over 12\pi^2}\int_\La^1 d\al~(m_b^2-s\al).
\enq
\beqa
\rho^{\qq\lag G^2\rag}(s)=0,
\enqa
\beqa
&&\rho^{\lag\bar{s}s\rag\lag G^2\rag}(s)=\frac{\lag\bar{s}s\rag\Gd}{2^83^2\pi^4}\bigg[-\int_0^1d\al\frac{m_b^3\alpha^2}{(1-\alpha)^3 }\times\nonumber\\&&+\int_\Lambda^1d\al\bigg(\frac{(1-\al)(12\al-3)m_b}{\al^2}-\frac{8m_b}{2}\bigg)\bigg],
\enqa

\beqa
\rho^{\lag G^4\rag}(s)&=&\frac{\lag g^4G^4\rag}{2^{13}3^2\pi^6}\bigg[\frac{3}{2}\int_\Lambda^1d\al-\int_0^1d\al\frac{m_b^2\al}{(1-\al)^2}\times\nonumber\\&&\times\delta\left(s-\frac{m_b^2}{1-\al}\right)\bigg].
\enqa

From $j_{ss}$ we get:
\beq
\rh^{m_s}(s)=-{m_sm_b\over 2^{8} 3\pi^6}\int_\La^1 d\al\left({1-\al\over\al}
\right)^3(m_b^2-s\al)^3,\label{rhoms}
\enq
\beqa
\rho^{\qq}(s)={1\over 2^{6}\pi^4}\int_\La^1 d\al{1-\al\over\al}(m_b^2-s\al)^2
\bigg[
\nn\\
\lag\bar{s}s\rag\left(2m_s-m_b{1-\al\over\al}\right)-2m_s\qq\bigg],
\enqa
\beqa
\rho^{mix}(s)&=&{1\over 2^{6}\pi^4}\int_\La^1 d\al~(m_b^2-s\al)\bigg[{\mixs
\over2}\bigg({m_s\over3}
\nn\\
&-&m_s{1-\al\over\al}
-m_b{1-\al\over\al}\left(1-{1-\al\over2\al}\right)\bigg)
\nn\\
&-&m_s\mix(1-\ln(1-\al))\bigg],
\enqa
\beqa
&&\rho^{\qq^2}(s)={\lag\bar{s}s\rag\over 48\pi^2}\left[\int_\La^1 d\al~\Big(4\qq(s\al-m_b^2)+2m_sm_b\times\right.\nonumber\\&&\left.\times(2\qq-\lag\bar{s}s\rag)\Big)+ms^2\Big(\lag\bar{s}s\rag-2\qq\Big)\right],
\enqa

\beqa
\rho^{\qq\lag G^2\rag}(s)&&=\frac{m_s\qq\Gd}{2^83^2\pi^4} \bigg[\int_0^1d\al\frac{2m_b^2\alpha}{(1-\al)^2}\left(1+\right.\nonumber\\&&\left.+\frac{m_bm_s}{(1-\al) M^2}\right)\delta\left(s-\frac{m_b^2}{(1-\al)}\right)-3\int_\Lambda^1d\al+\nonumber\\&&-(1+m_bm_s\,\delta(s-m_b^2))\bigg],
\enqa

\beqa
&&\rho^{\lag\bar{s}s\rag\lag G^2\rag}(s)=\frac{\lag\bar{s}s\rag\Gd}{2^83^2\pi^4}\bigg[\frac{3m_s}{2}-\int_0^1d\al\frac{m_b^2\alpha}{(1-\alpha)^2 }\times\nonumber\\&&\times\left(\frac{(\alpha\,M^2+m_s^2)m_b }{(1-\al)M^2}+2m_s \right) \delta\left(s-\frac{m_b^2}{1-\al}\right)\nonumber\\&&+\int_\Lambda^1d\al\bigg(\frac{3(1-\al)(4\al-1)m_b}{\al^2}+\frac{3m_s-8m_b}{2}\bigg)\bigg],
\enqa

\beqa
&&\rho^{\qq\mixs}(s)=m_bm_s\frac{\qq\mixs}{2^33^2\pi^2}\bigg[\int_0^1d\al\frac{3}{2(1-\al)}\times\nonumber\\&&\times\delta\left(s-\frac{m_b^2}{1-\al}\right)+(m_bm_s-3)\delta(s-m_b^2)\bigg)-\frac{3}{2}\bigg],
\enqa

\beqa
\rho^{\lag\bar{s}s\rag\mix}(s)&=&\frac{\lag\bar{s}s\rag\mix}{2^43\pi^2}\bigg[\frac{m_bm_s}{M^2}(m_bm_s-2M^2)\times\nonumber\\&&\times\delta(s-m_b^2)-1\bigg],
\enqa

\beqa
&&\rho^{\lag\bar{s}s\rag\mixs}(s)=m_bm_s\frac{\lag\bar{s}s\rag\mixs}{2^53^2\pi^2}\bigg[\frac{2M^2-m_bm_s}{M^2}\nonumber\\&&\times\delta(s-m_b^2)-\int_0^1d\al\frac{3}{1-\al}\delta\left(s-\frac{m_b^2}{1-\al}\right)\bigg],
\enqa

\beqa
\rho^{\lag G^4\rag}(s)&=&\frac{\lag g^4G^4\rag}{2^{13}3^2\pi^6}\bigg[\frac{3}{2}\int_\Lambda^1d\al-\int_0^1d\al\frac{m_b^2\al}{(1-\al)^2}\left(1+\nonumber\right.\\&&\left.+\frac{m_bm_s}{(1-\al)M^2}\right)\delta\left(s-\frac{m_b^2}{1-\al}\right)\bigg].
\enqa

\end{document}